\documentclass[12pt]{article}
\usepackage{amssymb}

\input{tcilatex}

\begin{document}

\bigskip\ 

\bigskip\ 

\begin{center}
\textbf{TOWARDS} \textbf{AN ASHTEKAR FORMALISM IN}

\smallskip\ 

\textbf{TWELVE DIMENSIONS}\footnote{%
Dedicated to Octavio Obreg\'{o}n on the occasion of his sixtieth birthday}

\textbf{\ }

\textbf{\ }

\smallskip\ 

J. A. Nieto\footnote{%
nieto@uas.uasnet.mx}

\smallskip

\textit{Facultad de Ciencias F\'{\i}sico-Matem\'{a}ticas de la Universidad
Aut\'{o}noma}

\textit{de Sinaloa, 80010, Culiac\'{a}n Sinaloa, M\'{e}xico}

\bigskip\ 

\bigskip\ 

\textbf{Abstract}
\end{center}

We discuss the Ashtekar formalism from the point of view of twelve
dimensions. We first focus on the $2+10$ spacetime signature and then we
consider the transition $2+10\rightarrow (2+2)+(0+8)$. We argue that both
sectors $2+2$ and $0+8,$ which are exceptional signatures, can be analyzed
from the point of view of a self-dual action associated with the Ashtekar
formalism.

\bigskip\ 

\bigskip\ 

\bigskip\ 

\bigskip\ 

\bigskip\ 


Pacs numbers: 04.60.-m, 04.65.+e, 11.15.-q, 11.30.Ly

October, 2006

\newpage

Over the years it has become clear that considering gravitational and gauge
theories with more than one time coordinate is a very interesting and useful
idea for understanding different aspects of traditional gravitational and
gauge field theories with only one time (see Ref. [1] and references
therein). In particular, the $2+10$-dimensional spacetime signature has
emerged as an interesting possibility for the understanding of both
supergravity and super Yang-Mills theory in $D=11$ (see Ref. [2]). Thus, by
seriously taking a $2+10$-dimensional gravity one may be interested in
various possibilities offered by this theory. For instance, one may be
interested in a realistic theory in four dimensions via the symmetry braking 
$2+10\rightarrow (1+3)+(1+7)$. However, this may not be the only attractive
possibility. In fact, one may think on the alternative transition

\begin{equation}
2+10\rightarrow (2+2)+(0+8).  \tag{1}
\end{equation}%
Of course, in this case one should not have a direct connection with our
four dimensional real World. Nevertheless, the signautre $2+2$ has been
considered in connection with a world volume of $2+2$-brane (see [3]). In
fact, the $2+2$-brane arises in $N=2$ theories which require two times for
its complete formulation [4]. Another source of interest in the $2+2$
signature is that such a signature admits a Majorana-Weyl spinors and
self-dual gauge fields formulation [5]. Moreover, it has been shown that the
symmetry $SL(2,R)$ makes the $2+2$ signature an exceptional one [6]. On the
other hand, the signature $0+8$ is euclidean and in principle can be treated
with the traditional methods such as the octonion algebraic approach [7]. In
pass, it is interesting to observe that octonion algebra is also exceptional
in the sense of Hurwitz theorem (see Ref. [8] and references therein). Thus,
we see that both $2+2$ and $0+8$ are exceptional signatures and therefore
these observations make the transition (1) worthwhile of being studied.

Here, we shall discuss the signatures $2+2$ and $0+8$ from the point of view
of `self-dual' actions associated with the Ashtekar formalism (see Ref. [9]
and references therein). For that purpose let us assume that the spacetime
manifold $M^{2+10}$ can be broken up into the form $M^{10+2}\rightarrow
M^{2+2}\times M^{0+8}.$ This implies that the $SO(2,10)$ Lovelock type
curvature (see Refs. [10]-[13] and references therein)

\begin{equation}
\mathcal{R}_{\hat{\mu}\hat{\nu}}^{\hat{A}\hat{B}}=R_{\hat{\mu}\hat{\nu}}^{%
\hat{A}\hat{B}}+\Sigma _{\hat{\mu}\hat{\nu},}^{\hat{A}\hat{B}}  \tag{2}
\end{equation}%
with\ 

\begin{equation}
R_{\hat{\mu}\hat{\nu}}^{\hat{A}\hat{B}}=\partial _{\hat{\mu}}\omega _{\hat{%
\nu}}^{\hat{A}\hat{B}}-\partial _{\hat{\nu}}\omega _{\hat{\mu}}^{\hat{A}\hat{%
B}}+\omega _{\hat{\mu}}^{\hat{A}\hat{C}}\omega _{\hat{\nu}\hat{C}}^{\hat{B}%
}-\omega _{\hat{\mu}}^{\hat{B}\hat{C}}\omega _{\hat{\nu}\hat{C}}^{\hat{A}} 
\tag{3}
\end{equation}%
and

\begin{equation}
\Sigma _{\hat{\mu}\hat{\nu}}^{\hat{A}\hat{B}}=e_{\hat{\mu}}^{\hat{A}}e_{\hat{%
\nu}}^{\hat{B}}-e_{\hat{\mu}}^{\hat{B}}e_{\hat{\nu}}^{\hat{A}},  \tag{4}
\end{equation}%
can be split into the form

\begin{equation}
\mathcal{R}_{ij}^{AB}=R_{ij}^{AB}+\Sigma _{ij}^{AB}  \tag{5}
\end{equation}%
and

\begin{equation}
\mathcal{R}_{\mu \nu }^{\hat{a}\hat{b}}=R_{\mu \nu }^{\hat{a}\hat{b}}+\Sigma
_{\mu \nu }^{\hat{a}\hat{b}},  \tag{6}
\end{equation}%
with the corresponding definitions (3) and (4) for $R_{ij}^{AB},\Sigma
_{ij}^{AB},R_{\mu \nu }^{\hat{a}\hat{b}}$ and $\Sigma _{\mu \nu }^{\hat{a}%
\hat{b}}$. In addition, we assume that $\mathcal{R}_{ij}^{AB}$ and $\mathcal{%
R}_{\mu \nu }^{\hat{a}\hat{b}}$ `live' in $M^{2+2}$ and $M^{0+8},$ with $%
SO(2,2)$ and $SO(8)$ as the corresponding gauge groups, respectively.

Let us now consider a MacDowell-Mansouri type action [11]-[13] in a $2+10-$%
dimensional spacetime [14]

\begin{equation}
\mathcal{S}=\int_{M^{2+10}}\Omega ^{\hat{\mu}\hat{\nu}\hat{\alpha}\hat{\beta}%
}\mathcal{R}_{\hat{\mu}\hat{\nu}}^{\hat{A}\hat{B}}\mathcal{R}_{\hat{\alpha}%
\hat{\beta}}^{\hat{C}\hat{D}}\Omega _{\hat{A}\hat{B}\hat{C}\hat{D}},  \tag{7}
\end{equation}%
where $\Omega ^{\hat{\mu}\hat{\nu}\hat{\alpha}\hat{\beta}}$ is a completely
antisymmetric constant in $M^{10+2}$ and $\Omega _{ABCD}$ is also a
completely antisymmetric constant associated with the $SO(2,10)$ group, yet
to be chosen. Assuming the transition $M^{10+2}\rightarrow M^{2+2}\times
M^{0+8}$ we find that the action (7) may be split as

\begin{equation}
\mathcal{S}=\int_{M^{2+2}}\varepsilon ^{ijkl}\mathcal{R}_{ij}^{AB}\mathcal{R}%
_{kl}^{CD}\varepsilon _{ABCD}+\int_{M^{0+8}}\eta ^{\mu \nu \alpha \beta }%
\mathcal{R}_{\mu \nu }^{\hat{a}\hat{b}}\mathcal{R}_{\alpha \beta }^{\hat{c}%
\hat{d}}\eta _{\hat{a}\hat{b}\hat{c}\hat{d}}.  \tag{8}
\end{equation}%
Here, $\varepsilon ^{ijkl}$ and $\varepsilon _{ABCD}$ are completely
antisymmetric objects linked to the signature $2+2$, while $\eta ^{\mu \nu
\alpha \beta }$ and $\eta _{\hat{a}\hat{b}\hat{c}\hat{d}}$ are completely
antisymmetric objects linked to the signature $0+8$. The next step in our
quest of associating an Ashtekar formalism with the spacetimes of signatures 
$2+2$ and $0+8$ is to consider the self-dual (antiself-dual) sector of the
action (8).

Let us first focus on the first term in (8);

\begin{equation}
S_{2+2}=\int_{M^{2+2}}\varepsilon ^{ijkl}\mathcal{R}_{ij}^{AB}\mathcal{R}%
_{kl}^{CD}\varepsilon _{ABCD}.  \tag{9}
\end{equation}%
We recognize this action as the MacDowell-Mansouri action for spacetime of
signature $2+2$. Before we write the self dual sector of (9) it is
convenient to discuss some of the properties of the object $\varepsilon
_{ABCD}$. First, let us set $\varepsilon _{1234}=1$. So we find that

\begin{equation}
\varepsilon ^{ABCD}=\eta ^{AE}\eta ^{BF}\eta ^{CG}\eta ^{DH}\varepsilon
_{EFGH},  \tag{10}
\end{equation}%
where $\eta _{AB}=diag(-1,-1,1,1)$, gives $\varepsilon ^{1234}=1$. We
observe that $Det\eta _{AB}=1$. Hence, we get

\begin{equation}
\varepsilon ^{ABCD}\varepsilon _{EFCD}=2\delta _{EF}^{AB}.  \tag{11}
\end{equation}%
Here, we used the definition $\delta _{EF}^{AB}\equiv \delta _{E}^{A}\delta
_{F}^{B}-\delta _{E}^{B}\delta _{F}^{A}$, where $\delta _{E}^{A}$ is the
Kronecker delta. This means that the property (11) of $\varepsilon _{ABCD}$
is exactly the the same as the corresponding euclidean quantity.

Let us now define the dual curvature

\begin{equation}
^{\ast }\mathcal{R}_{ij}^{AB}=\frac{1}{2}\varepsilon _{CD}^{AB}\mathcal{R}%
_{ij}^{CD}.  \tag{12}
\end{equation}%
Using (11) we see that

\begin{equation}
^{\ast \ast }\mathcal{R}_{ij}^{AB}=\mathcal{R}_{ij}^{AB}.  \tag{13}
\end{equation}%
In this way, the self-dual (antiself-dual) curvature

\begin{equation}
^{\pm }\mathcal{R}_{ij}^{AB}=\frac{1}{2}(\mathcal{R}_{ij}^{AB}\pm ^{\ast }%
\mathcal{R}_{ij}^{AB})  \tag{14}
\end{equation}%
gives

\begin{equation}
^{\ast \pm }\mathcal{R}_{ij}^{AB}=(\pm )^{\pm }\mathcal{R}_{ij}^{AB}, 
\tag{15}
\end{equation}%
which means that $^{\pm }\mathcal{R}_{ij}^{AB}$ is self dual
(antiself-dual). In fact, we have

\[
\mathcal{R}_{ij}^{AB}=^{+}\mathcal{R}_{ij}^{AB}+^{-}\mathcal{R}_{ij}^{AB}. 
\]%
This implies that the action (see Refs. [13], [15], [16])

\begin{equation}
^{+}\mathcal{S}_{2+2}=\int_{M^{2+2}}\varepsilon ^{ijkl+}\mathcal{R}%
_{ij}^{AB+}\mathcal{R}_{kl}^{CD}\varepsilon _{ABCD}  \tag{16}
\end{equation}%
corresponds to the self-dual sector of the action (9). A similar action can
be written for the antiself-dual sector of (9). In fact, we have $\mathcal{S}%
_{2+2}=^{+}\mathcal{S}_{2+2}+^{-}\mathcal{S}_{2+2}$.

Now we would like to discuss the consequences of (16). For this purpose we
first write (14) in the form

\begin{equation}
^{\pm }\mathcal{R}_{ij}^{AB}=\frac{1}{2}\ ^{\pm }B_{KL}^{AB}\mathcal{R}%
_{ij}^{KL},  \tag{17}
\end{equation}%
where

\begin{equation}
^{\pm }B_{KL}^{AB}=\frac{1}{2}(\delta _{KL}^{AB}\pm \varepsilon _{KL}^{AB}).
\tag{18}
\end{equation}%
By straightforward computation one finds that the projector $^{\pm
}B_{KL}^{AB}$ satisfies the property

\begin{equation}
^{\pm }B_{MN}^{AB\pm }B_{RS}^{CD}\varepsilon _{ABCD}=\pm 4^{\pm }B_{MNRS}. 
\tag{19}
\end{equation}%
Therefore, using (19) we see that (16) can also be written as

\begin{equation}
^{+}S_{2+2}=\int_{M^{2+2}}\varepsilon ^{ijkl}\mathcal{R}_{ij}^{AB}\mathcal{R}%
_{kl}^{CD}B_{ABCD}.  \tag{20}
\end{equation}%
Using (18) we discover that (20) results in

\begin{equation}
^{+}\mathcal{S}_{2+2}=\frac{1}{2}\int_{M^{2+2}}\varepsilon ^{ijkl}\mathcal{R}%
_{ij}^{AB}\mathcal{R}_{kl}^{CD}\varepsilon _{ABCD}+\frac{1}{2}%
\int_{M^{2+2}}\varepsilon ^{ijkl}\mathcal{R}_{ij}^{AB}\mathcal{R}%
_{kl}^{CD}\eta _{ABCD},  \tag{21}
\end{equation}%
where $\eta _{ABCD}\equiv \eta _{AC}\eta _{BD}-\eta _{AD}\eta _{BC}$. The
first term in (21) can be split using (5) in the Euler topological invariant
and the Einstein-Hilbert action with cosmological constant, while the second
term leads to the Pontrjagin topological invariant. Therefore, up to
topological invariants the action (16) is equivalent to the Einstein-Hilbert
action with cosmological constant. It is worth mentioning how the
cosmological constant arises from the first term of (21). Since $\mathcal{R}%
_{ij}^{AB}=R_{ij}^{AB}+\Sigma _{ij}^{AB}$ and $\Sigma
_{ij}^{AB}=e_{i}^{A}e_{j}^{B}-e_{i}^{B}e_{j}^{A}$ we observe that under the
rescaling $e_{i}^{A}\rightarrow \lambda e_{i}^{A}$ we shall get the
transformations $\varepsilon ^{ijkl}\Sigma _{ij}^{AB}R_{kl}^{CD}\varepsilon
_{ABCD}\rightarrow \lambda ^{2}\varepsilon ^{ijkl}\Sigma
_{ij}^{AB}R_{kl}^{CD}\varepsilon _{ABCD}$ and $\varepsilon ^{ijkl}\Sigma
_{ij}^{AB}\Sigma _{kl}^{CD}\varepsilon _{ABCD}\rightarrow \lambda
^{4}\varepsilon ^{ijkl}\Sigma _{ij}^{AB}\Sigma _{kl}^{CD}\varepsilon _{ABCD}$
which means that under the rescaling $^{+}\mathcal{S}_{2+2}\rightarrow \frac{%
1}{\lambda ^{2}}^{+}\mathcal{S}_{2+2}$ one can identify, up to numerical
factor, $\lambda ^{2}$ with the cosmological constant (see Ref. [12] for
details).

Let us now consider the second term in (8) (see Ref. [14])

\begin{equation}
S_{0+8}=\int_{M^{0+8}}\eta ^{\mu \nu \alpha \beta }\mathcal{R}_{\mu \nu }^{%
\hat{a}\hat{b}}\mathcal{R}_{\alpha \beta }^{\hat{c}\hat{d}}\eta _{\hat{a}%
\hat{b}\hat{c}\hat{d}}.  \tag{22}
\end{equation}%
First we need to clarify the meaning of the completely antisymmetric objects 
$\eta _{\hat{a}\hat{b}\hat{c}\hat{d}}$ (or $\eta ^{\mu \nu \alpha \beta }$).
The key idea is to relate $\eta _{\hat{a}\hat{b}\hat{c}\hat{d}}$ to the
octonion structure constants $C_{\hat{a}\hat{b}}^{\hat{c}}$ in the form

\begin{equation}
\eta _{8abc}\equiv \varsigma C_{abc}  \tag{23}
\end{equation}%
and

\begin{equation}
\eta _{abcd}\equiv \frac{1}{3!}\varepsilon _{abcdefg}C^{efg},  \tag{24}
\end{equation}%
where the indices $a,b,...etc$ run from $1$ to $7$, $\varepsilon _{abcdefg}$
is the completely antisymmetric symbol in seven dimensions and $\varsigma
=\pm $. Using (23) and (24) it can be shown that $\eta _{\hat{a}\hat{b}\hat{c%
}\hat{d}}$ is self-dual:

\begin{equation}
\eta _{\hat{a}\hat{b}\hat{c}\hat{d}}=\frac{\varsigma }{4!}\varepsilon _{\hat{%
a}\hat{b}\hat{c}\hat{d}\hat{e}\hat{f}\hat{g}\hat{h}}\eta ^{\hat{e}\hat{f}%
\hat{g}\hat{h}}.  \tag{25}
\end{equation}%
For $\varsigma =1,$ it is self-dual (and for $\varsigma =-1$ is
antiself-dual). One can verify that four-rank completely antisymmetric
tensor $\eta _{\hat{a}\hat{b}\hat{c}\hat{d}}$ (also $\eta _{\mu \nu \alpha
\beta }$) satisfies the relations [17]-[19] (see also Refs. [7] and [20]),

\begin{equation}
\eta _{\hat{a}\hat{b}\hat{c}\hat{d}}\eta ^{\hat{e}\hat{f}\hat{c}\hat{d}%
}=6\delta _{\hat{a}\hat{b}}^{\hat{e}\hat{f}}+4\eta _{\hat{a}\hat{b}}^{\hat{e}%
\hat{f}},  \tag{26}
\end{equation}%
\begin{equation}
\eta _{\hat{a}\hat{b}\hat{c}\hat{d}}\eta ^{\hat{e}\hat{b}\hat{c}\hat{d}%
}=42\delta _{\hat{a}}^{\hat{e}},  \tag{27}
\end{equation}%
\begin{equation}
\eta _{\hat{a}\hat{b}\hat{c}\hat{d}}\eta ^{\hat{a}\hat{b}\hat{c}\hat{d}}=336.
\tag{28}
\end{equation}

The next step is to introduce the dual of $\mathcal{R}_{\mu \nu }^{\hat{a}%
\hat{b}}$ in the form

\begin{equation}
^{\star }\mathcal{R}_{\mu \nu }^{\hat{a}\hat{b}}=\frac{1}{2}\eta _{\hat{c}%
\hat{d}}^{\hat{a}\hat{b}}\mathcal{R}_{\mu \nu }^{\hat{c}\hat{d}}.  \tag{29}
\end{equation}%
The self-dual and antiself-dual parts $^{\pm }\mathcal{R}_{\mu \nu }^{\hat{a}%
\hat{b}}$ of $\mathcal{R}_{\mu \nu }^{\hat{a}\hat{b}}$ are defined as

\begin{equation}
^{+}\mathcal{R}_{\mu \nu }^{\hat{a}\hat{b}}=\frac{1}{4}(\mathcal{R}_{\mu \nu
}^{\hat{a}\hat{b}}+^{\star }\mathcal{R}_{\mu \nu }^{\hat{a}\hat{b}}) 
\tag{30}
\end{equation}%
and%
\begin{equation}
^{-}\mathcal{R}_{\mu \nu }^{\hat{a}\hat{b}}=\frac{1}{4}(3\mathcal{R}_{\mu
\nu }^{\hat{a}\hat{b}}-^{\star }\mathcal{R}_{\mu \nu }^{\hat{a}\hat{b}}), 
\tag{31}
\end{equation}%
respectively. Since

\begin{equation}
^{\star \star }\mathcal{R}_{\mu \nu }^{\hat{a}\hat{b}}=3\mathcal{R}_{\mu \nu
}^{\hat{a}\hat{b}}+2^{\star }\mathcal{R}_{\mu \nu }^{\hat{a}\hat{b}}, 
\tag{32}
\end{equation}%
we see that

\begin{equation}
^{\star +}\mathcal{R}_{\mu \nu }^{\hat{a}\hat{b}}=3^{+}\mathcal{R}_{\mu \nu
}^{\hat{a}\hat{b}}  \tag{33}
\end{equation}%
and

\begin{equation}
^{\star -}\mathcal{R}_{\mu \nu }^{\hat{a}\hat{b}}=-^{-}\mathcal{R}_{\mu \nu
}^{\hat{a}\hat{b}}.  \tag{34}
\end{equation}%
Thus, up to a numerical factor we see that $^{+}\mathcal{R}_{\mu \nu }^{\hat{%
a}\hat{b}}$ and $^{-}\mathcal{R}_{\mu \nu }^{\hat{a}\hat{b}}$ play, in fact,
the role of the self-dual and antiself-dual parts, respectively of $\mathcal{%
R}_{\mu \nu }^{\hat{a}\hat{b}}.$ It turns out to be convenient to write (30)
as

\begin{equation}
^{+}\mathcal{R}_{\mu \nu }^{\hat{a}\hat{b}}=\frac{1}{2}\ ^{+}\Lambda _{\hat{c%
}\hat{d}}^{\hat{a}\hat{b}}\mathcal{R}_{\mu \nu }^{\hat{c}\hat{d}},  \tag{35}
\end{equation}%
where

\begin{equation}
^{+}\Lambda _{\hat{c}\hat{d}}^{\hat{a}\hat{b}}=\frac{1}{4}(\delta _{\hat{c}%
\hat{d}}^{\hat{a}\hat{b}}+\eta _{\hat{c}\hat{d}}^{\hat{a}\hat{b}}).  \tag{36}
\end{equation}%
While, (31) can be written in the form

\begin{equation}
^{-}\mathcal{R}_{\mu \nu }^{\hat{a}\hat{b}}=\frac{1}{2}\ ^{-}\Lambda _{\hat{c%
}\hat{d}}^{\hat{a}\hat{b}}\mathcal{R}_{\mu \nu }^{\hat{c}\hat{d}},  \tag{37}
\end{equation}%
with

\begin{equation}
^{-}\Lambda _{\hat{c}\hat{d}}^{\hat{a}\hat{b}}=\frac{1}{4}(3\delta _{\hat{c}%
\hat{d}}^{\hat{a}\hat{b}}-\eta _{\hat{c}\hat{d}}^{\hat{a}\hat{b}}).  \tag{38}
\end{equation}

Now, we would like to propose the action

\begin{equation}
^{\pm }\mathcal{S}_{0+8}=\frac{1}{^{+}\tau }\int_{M^{0+8}}\eta ^{\mu \nu
\alpha \beta +}\mathcal{R}_{\mu \nu }^{\hat{a}\hat{b}+}\mathcal{R}_{\alpha
\beta }^{\hat{c}\hat{d}}\eta _{\hat{a}\hat{b}\hat{c}\hat{d}}+\frac{1}{%
^{-}\tau }\int_{M^{0+8}}\eta ^{\mu \nu \alpha \beta -}\mathcal{R}_{\mu \nu
}^{\hat{a}\hat{b}-}\mathcal{R}_{\alpha \beta }^{\hat{c}\hat{d}}\eta _{\hat{a}%
\hat{b}\hat{c}\hat{d}},  \tag{39}
\end{equation}%
which is a generalization of the action (22). Here, $^{+}\tau $ and $%
^{-}\tau $ are two constant parameters. It is worth mentioning that the
proposal (39) emerged from the observation that $^{\pm }\Lambda $ are
projection operators. In fact, one can prove that the objects $^{+}\Lambda $
and $^{-}\Lambda ,$ given in (36) and (38) respectively, satisfy [19]

\begin{equation}
^{+}\Lambda +^{-}\Lambda =1,  \tag{40}
\end{equation}

\begin{equation}
^{+}\Lambda ^{-}\Lambda =^{-}\Lambda ^{+}\Lambda =0,  \tag{41}
\end{equation}

\begin{equation}
^{+}\Lambda ^{2}=^{+}\Lambda ,  \tag{42}
\end{equation}%
and

\begin{equation}
^{-}\Lambda ^{2}=^{-}\Lambda .  \tag{43}
\end{equation}%
Here, $^{\pm }\Lambda ^{2}$ mean $\frac{1}{4}^{\pm }\Lambda _{\hat{c}\hat{d}%
}^{\hat{a}\hat{b}\pm }\Lambda _{\hat{g}\hat{h}}^{\hat{e}\hat{f}}\delta _{%
\hat{a}\hat{b}\hat{e}\hat{f}}$.

Let us focus on the self-dual part of (39):

\begin{equation}
^{+}\mathcal{S}_{0+8}=\frac{1}{^{+}\tau }\int_{M^{0+8}}\eta ^{\mu \nu \alpha
\beta +}\mathcal{R}_{\mu \nu }^{\hat{a}\hat{b}+}\mathcal{R}_{\alpha \beta }^{%
\hat{c}\hat{d}}\eta _{\hat{a}\hat{b}\hat{c}\hat{d}}.  \tag{44}
\end{equation}%
Presumably, most of the computations that we shall develop below in
connection with $^{+}\mathcal{S}_{0+8}$ may also be applied to the
antiself-dual sector $^{-}\mathcal{S}_{0+8}.$ It is worth mentioning that
the action (39) is the analogue of the action proposed by Nieto [14] in
eight dimensions with signature $1+7$. Let us start observing that since

\begin{equation}
^{+}\mathcal{R}_{\mu \nu }^{\hat{a}\hat{b}}=^{+}R_{\mu \nu }^{\hat{a}\hat{b}%
}+^{+}\Sigma _{\mu \nu }^{\hat{a}\hat{b}},  \tag{45}
\end{equation}%
one finds that the action (44) becomes

\begin{equation}
^{+}\mathcal{S}_{0+8}=\frac{1}{^{+}\tau }\int_{M^{0+8}}(T+K+C),  \tag{46}
\end{equation}%
with

\begin{equation}
T=\eta ^{\mu \nu \alpha \beta +}R_{\mu \nu }^{\hat{a}\hat{b}+}R_{\alpha
\beta }^{\hat{c}\hat{d}}\eta _{\hat{a}\hat{b}\hat{c}\hat{d}},  \tag{47}
\end{equation}

\begin{equation}
K=2\eta ^{\mu \nu \alpha \beta +}\Sigma _{\mu \nu }^{\hat{a}\hat{b}%
+}R_{\alpha \beta }^{\hat{c}\hat{d}}\eta _{\hat{a}\hat{b}\hat{c}\hat{d}}, 
\tag{48}
\end{equation}%
and

\begin{equation}
C=\eta ^{\mu \nu \alpha \beta +}\Sigma _{\mu \nu }^{\hat{a}\hat{b}+}\Sigma
_{\alpha \beta }^{\hat{c}\hat{d}}\eta _{\hat{a}\hat{b}\hat{c}\hat{d}}. 
\tag{49}
\end{equation}%
Using (35) and (36), it is not difficult to see that $T$ can be identified
with a topological invariant in eight dimensions analogous to Pontrjagin and
Euler invariants in four dimensions. At this respect, it is worth mentioning
that in the case of $G_{2}$-invariant super Yang Mills theory [21] a
topological term of the form

\begin{equation}
\mathcal{S}_{0+8}=\frac{1}{^{+}\tau }\int_{M^{0+8}}\eta ^{\mu \nu \alpha
\beta }F_{\mu \nu }^{a}F_{\alpha \beta }^{b}g_{ab},  \tag{50}
\end{equation}%
where $F_{\mu \nu }^{a}$ is the Yang-Mills field strength and $g_{ab}$ is
the group invariant metric, has been considered. Thus, the term $T$ in (47)
can be considered as the `gravitational' analogue of (50). Similarly, $K$
should lead to a kind of gravity in eight dimensions. Finally, $C$ may be
identified as the analogue of a cosmological constant term in the following
sense. For a cosmological constant one would expect a term of the form

\begin{equation}
\frac{1}{8!}\int_{M^{0+8}}\varepsilon ^{\mu _{1}...\mu _{8}}\varepsilon _{%
\hat{a}_{1}...\hat{a}_{8}}e_{\mu _{1}}^{\hat{a}_{1}}\cdot \cdot \cdot e_{\mu
_{8}}^{\hat{a}_{8}}=\int_{M^{0+8}}\det (e_{\mu }^{\hat{a}}).  \tag{51}
\end{equation}%
But, $C$ is quartic in $e_{\mu }^{\hat{a}}$ and then a first sight one may
say that does not contain $\det (e_{\mu }^{\hat{a}}).$ However, due to the
self-dual relation (25) one may write $C$ in the form

\begin{equation}
C=\frac{1}{4!}\eta ^{\mu \nu \alpha \beta +}\Sigma _{\mu \nu }^{\hat{a}\hat{b%
}+}\Sigma _{\alpha \beta }^{\hat{c}\hat{d}}\varepsilon _{\hat{a}\hat{b}\hat{c%
}\hat{d}\hat{e}\hat{f}\hat{g}\hat{h}}\eta ^{\hat{e}\hat{f}\hat{g}\hat{h}}. 
\tag{52}
\end{equation}%
Thus, using the identity $\varepsilon _{\hat{a}_{1}...\hat{a}_{8}}e_{\mu
_{1}}^{\hat{a}_{1}}\cdot \cdot \cdot e_{\mu _{8}}^{\hat{a}_{8}}=\det (e_{\mu
}^{\hat{a}})\varepsilon _{\mu _{1}...\mu _{8}}$ one may obtain that $C\sim
\det (e_{\mu }^{\hat{a}})$. In fact, (52) can be rewritten as

\begin{equation}
\begin{array}{c}
C=\frac{1}{4!}\eta ^{\mu \nu \alpha \beta +}\Sigma _{\mu \nu }^{\tau \lambda
+}\Sigma _{\alpha \beta }^{\sigma \rho }\varepsilon _{\hat{a}\hat{b}\hat{c}%
\hat{d}\hat{e}\hat{f}\hat{g}\hat{h}}e_{\tau }^{\hat{a}}e_{\lambda }^{\hat{b}%
}e_{\sigma }^{\hat{c}}e_{\rho }^{\hat{d}}e_{\gamma }^{\hat{e}}e_{\delta }^{%
\hat{f}}e_{\xi }^{\hat{g}}e_{\theta }^{\hat{h}}\tilde{\eta}^{\gamma \delta
\xi \theta } \\ 
\\ 
=\det (e_{\mu }^{\hat{a}})\frac{1}{4!}\eta ^{\mu \nu \alpha \beta +}\Sigma
_{\mu \nu }^{\tau \lambda +}\Sigma _{\alpha \beta }^{\sigma \rho
}\varepsilon _{\tau \lambda \sigma \rho \gamma \delta \xi \theta }\tilde{\eta%
}^{\gamma \delta \xi \theta },%
\end{array}
\tag{53}
\end{equation}%
where $^{+}\Sigma _{\mu \nu }^{\tau \lambda }e_{\tau }^{\hat{a}}e_{\lambda
}^{\hat{b}}=^{+}\Sigma _{\mu \nu }^{\hat{a}\hat{b}}$ and $e_{\gamma }^{\hat{e%
}}e_{\delta }^{\hat{f}}e_{\xi }^{\hat{g}}e_{\theta }^{\hat{h}}\tilde{\eta}%
^{\gamma \delta \xi \theta }=\eta ^{\hat{e}\hat{f}\hat{g}\hat{h}}$. This
shows that in principle we may have $C\sim \det (e_{\mu }^{\hat{a}}).$
However, in the process we have introduced two new objects $^{+}\Sigma _{\mu
\nu }^{\tau \lambda }$ and $\tilde{\eta}^{\gamma \delta \xi \theta }$ which
may lead to an alternative result when they are again written in terms of $%
^{+}\Sigma _{\mu \nu }^{\hat{a}\hat{b}}$ and $\eta ^{\hat{e}\hat{f}\hat{g}%
\hat{h}}$ respectively. Other possibility is to consider $C$ as a new type
of cosmological constant term not necessarily related to $\det (e_{\mu }^{%
\hat{a}})$. In this case it is necessary to consider that, in general, the $%
\varepsilon $-symbol is Lorentz invariant in any dimension, but in contrast
the $\eta $-symbol is only $SO(7)$-invariant in eight dimensions (see Ref.
[19]). Therefore, the $\eta $-symbol spoils the Lorentz invariance of the
action $^{+}\mathcal{S}_{0+8}$ given in (50) and in particular of the $C$
term, but maintains a hidden $SO(7)$-invariance. In fact, this is a general
phenomenon in field theories involving the $\eta $-symbol (see, for
instance, Refs. [21] and [7]).

Let us summarize our results. We started with a $2+10$ dimensional
gravitational theory and we assumed a possible symmetry braking of the form $%
2+10\rightarrow (2+2)+(0+8)$. We proved that classically it makes sense to
associate both signatures $2+2$ and $0+8$ with the Ashtekar formalism.
Although our procedure was similar to the case $2+10\rightarrow (1+3)+(1+7)$%
, the steps were necessary if eventually one desires to develop an Ashtekar
canonical quantization for the signatures $2+2$ and $0+8$.

Since one of the most interesting candidates for the so-called $M-$theory is
a theory of a $2+2-$brane embedded in $2+10$ dimensional background target
spacetime (see [3] and Refs. therein) our formalism points out a possible
connection between Ashtekar formalism and $M-$theory. In fact, this version
of $M-$theory evolved from the observation [4] that the complex structure of 
$N=2$ strings requires a target spacetime of signature $2+2$ rather than $%
1+9 $ as the usual $N=1$ string theory. Thus, a natural step forward was to
consider the $N=(2,1)$ heterotic string [22]. In this scenario, it was
observed that a consistent $N=(2,1)$ string should consider right-movers
`living' in $2+2$ dimensions and left-movers in $2+10$ dimensions. The
connection with our work comes from the fact that the dynamics of a $2+2-$%
branes leads to self-dual gravity coupled to self-dual supermatter in $2+10$
dimensions. And self-duality in this signatures is precisely what we have
considered in this work.

Another reason for expecting a connection between $M-$theory and Ashtekar
formalism comes from the link between oriented matroid theory [23] and two
time physics. In fact, it has been proved [24] that oriented matroids may be
related to $M-$theory by various routes [25]-[29], and in particular via two
time physics [24]. Moreover, it has been proposed that oriented matroid
theory may provide a mathematical framework for $M-$theory [29]-[30]. Thus,
a connection between Ashtekar formalism and oriented matroid theory seems to
be an interesting possibility.

The actions $^{+}\mathcal{S}_{2+2}$ and $^{+}\mathcal{S}_{0+8},$ given in
(16) and (44) respectively, should in principle admit steps toward a
canonical quantization similar to the steps given after the
Jacobson-Smolin-Samuel action in four dimensions [31]-[32]. However, while
canonical quantization in four dimensions of the Jacobson-Smolin-Samuel
action leads to quantum states of the form $\exp (S_{cs}),$ where $S_{cs}$
is a Chern-Simons action, the quantum states in the signatures $2+2$ and $%
0+8 $ could be very different and surprising. The main reason for this is
that the signatures $2+2$ and $0+8$ are exceptional, and therefore, one
should expect that the corresponding canonical quantizations are also
exceptional.

We should mention something about the transition $M^{10+2}\rightarrow
M^{2+2}\times M^{0+8}$ which allows us to obtain (8) from the action (7).
Typically, in Kaluza-Klein theories, one assumes the compactification $%
M^{d+1}\rightarrow M^{3+1}\times B$, where $M^{3+1}$ is identified with the
ordinary four dimensional manifold and $B$ is considered as $(d-3)$%
-dimensional compact manifold. For instance, in $D=11$ supergravity one may
consider the compactification $M^{10+1}\rightarrow M^{3+1}\times S^{7}$,
where $S^{7}$ is the seven sphere. It is assumed that the "size" of the
compact manifold $B,$ with isometry group $G$, is much smaller than the
"size" of the other physical manifold $M^{3+1}$. This assumptions allow to
apply the so called dimensional reduction mechanism. In going from (7) to
(8) we have applied similar procedure considering $M^{0+8}$ as a kind of
compact manifold and in this way eliminating possible cross terms. However,
just as the Kaluza-Klein theory has several interesting generalizations,
including arbitrary $B$ manifolds (see, for instance, Ref. [33]), for
further research it may be interesting to investigate a possible interaction
between the two sectors $M^{2+2}$ and $M^{0+8}$.

Finally, it is worth mentioning that, besides the $N=(2,1)$ heterotic string
theory, Matrix theory [34] is another proposed candidate for $M-$theory.
Since there seems to exist a connection between these two approaches (see
Ref. [35]) one should also expect a relation between Ashtekar formalism and
Matrix theory. Fortunately, Smolin [36]-[37] (see also Refs \ [38] and [39])
already has described this possibility. In particular, in the context of
topological $M-$theory Smolin [37] has investigated the possibility of
obtaining Hitchin's 7 seven dimensional theory, which in principle seems to
admit background independent formulation, from the classical limit of $M-$%
theory, namely eleven dimensional supergravity. The idea is focused on an
attempt of reducing the eleven dimensional manifold $M^{1+10}$ in the form

\begin{equation}
M^{1+10}\rightarrow R\times \Sigma \times S^{1}\times R^{3}.  \tag{54}
\end{equation}%
Here, $\Sigma $ is a complex six-dimensional manifold. Considering that the
only degree of freedom is the gauge field three form $A$ which is pure gauge 
$A=d\beta $ and therefore locally trivial $dA=0$, the Smolin's conjecture is
that the Hitchin's action can be derived from the lowest dimensional term
that can be made from $d\beta $ on $R\times \Sigma $ of the corresponding
effective action (see Ref. [37] for details). Observing that $\Sigma \times
S^{1}$ is a seven dimensional manifold and since, via the octonion
structure, the solution $0+8$ is related to the seven sphere solution of
eleven dimensional supergravity one is motivated to conjecture that there
must be a connection between our approach of incorporating Ashtekar
formalism in the context of $M$-theory and the Smolin's program.

\end{document}